\documentclass[prd,showpacs]{revtex4}
\usepackage{amsmath}
\usepackage{amsfonts}
\usepackage{amssymb}
\usepackage{latexsym}
\usepackage[dvips]{graphicx}
\def\prd{Phys.\ Rev.\ D}
\def\plb{Phys.\ Lett.\ B}
\def\mn{\mu\nu}

\def\S{{\cal S}} 
\def\prd{Phys. Rev. D}

\def\be{\begin{equation}}
\def\ee{\end{equation}}  
\def\ba{\begin{eqnarray}}
\def\ea{\end{eqnarray}}  

\begin{document}
%



\title{An effective action for asymptotically safe gravity} 

\author{Alfio Bonanno$^{1,2}$}

\affiliation{$^1$INAF, Osservatorio Astrofisico di Catania, Via S.Sofia 78, 95123 Catania, Italy\\
$^2$INFN, Sezione di Catania, Via S.Sofia 72, 95123 Catania, Italy}

\begin{abstract}
Asymptotically safe theories of gravitation have received great attention in recent times. 
In this framework an effective action embodying the basic features of the renormalized flow
around the non-gaussian fixed point is derived and its implications for the early 
universe are discussed.  In particular, a ``landscape" of a countably infinite number 
of cosmological inflationary solutions characterized by an unstable de Sitter phase lasting  
for a large enough number of e-folds is found.
\end{abstract}

\pacs{04.60.-m, 04.50.Kd, 11.10.Hi, 98.80.Cq}

\maketitle
Thanks to the application of Renormalization Group (RG) techniques \cite{weinberg}, 
the possibility that the high energy behavior of gravity can be governed by an ultraviolet non-Gaussian fixed point (NGFP) 
is gradually emerging as a viable theoretical scenario for a consistent theory of Quantum Gravity.
In this context, the non-perturbative Functional Renormalization Group Equation (FRGE) \cite{chw} for gravity \cite{1998PhRvD..57..971R},
is a particularly important tool because it
generates an RG flow on a theory space which consists of all diffeomorphism invariant
functionals of the metric $g_{\mu\nu}$ . 
It defines a one parameter family of effective field theories with actions 
$\Gamma_k(g_{\mu\nu})$ depending on a coarse graining scale (or ``cutoff'') $k$,
and interpolates between a ``bare'' action (for $k\to\infty$)
and the ordinary effective action (for $k\to 0$).
When applied to the Einstein-Hilbert (EH) action the FRGE yields beta functions
\cite{1998PhRvD..57..971R,1998CQGra..15.3449D} 
which have made detailed investigations of the scaling behavior of Newtons's constant possible
\cite{1999PThPh.102..181S,
2002PhRvD..65b5013L,
2002PhRvD..66b5026L,
2002CQGra..19..483L,
2002PhRvD..65f5016R,
2002PhRvD..66l5001R,
2004PhRvL..92t1301L,
2005JHEP...02..035B,
2009GReGr..41..983R,
2009PhRvD..79j5005R,
2009arXiv0904.2510M,
2009arXiv0905.4220M}.
It has been shown quite convincingly that the dimensionful Newton 
constant $G$ is {\em antiscreened} at high energies,
a behavior that eventually leads to an ultraviolet (UV) NGFP, a necessary condition for asymptotic safety.
These analyses have then been enlarged to include matter \cite{2003PhRvD..67h1503P} and 
a growing number of purely gravitational operators in the action
\cite{2006PhRvL..97v1301C,bms1,Niedermaier:2009zz, ms1,cpr2}
 and reviews of these works have appeared in \cite{reviews}.

Clearly, a natural arena for the applications of this idea is the physics of the early universe and the space-time singularities.
In those contexts, the RG flow of the effective average action, obtained by different truncations 
of theory space, has been the basis of various investigations of ``RG improved" black hole \cite{bh,bh2}
and cosmological \cite{cosmo1,cosmofrank,cosmo2,elo,esposito, rw1, rw2} spacetimes. 
In particular, very recently it has been shown that the ``RG improved'' Einstein equations admit (power-law or exponential) inflationary solutions
and that the running of the cosmological constant can account for the entire entropy of the present universe in the massless sector 
\cite{2007JCAP...08..024B,2008JPhCS.140a2008B}  (see \cite{claqg08} for an extended review.)

However, it is still not always clear how to extract all the relevant information encoded in the running of the Newton constant  
because of the conceptual difficulties in relating the typical energy momentum scale $k$ with the space-time properties. 
In a situation with more than one possible physical cutoff scale the general theory 
of the effective average action  \cite{ergrew}  implies that the 
relevant action is $\Gamma_k$ where $k$ is the largest one of the various competing scale. 
Therefore in a realistic description of the early universe, including the
presence of the matter fields, the cutoff identification becomes a difficult task unless 
one knows a priori the hierarchies of scales during the universe evolution. However, 
even if this information were known, the cutoff  must then be promoted to the status of
a dynamical variable in order to preserve diffeomorphism invariance at at the level of the action 
\cite{bocoper,hilira,cohira,zottone}.
%
In a recent work \cite{weinberginflation}, Weinberg has discussed the possibility  that, in the context of 
asymptotically safe gravity, it could be possible to have a period of exponential expansion which comes to an end after enough e-folds of inflation.  
An important ingredient is the choice of an ``optimal cut-off'' which only partially includes the effect of the radiative corrections.
Unfortunately, it turns out that an unstable de Sitter phase which lasts enough e-folds
is in general not simple to achieve in this model, unless some fine-tuning occurs \cite{weinberginflation,tye}. 

Ideally, if one wanted to study the quantum evolution of the curvature at very high, let's say planckian
energies, the full effective action $\Gamma(g_{\mu\nu})$ should in principle be used.
This functional coincides with $\Gamma_k(g_{\mu\nu})$ in the limit $k\to 0$ but this limit is 
not easily accessible unless renormalization effects are unimportant below some given mass scale.

A possible strategy in  order to obtain an effective action describing the physics near the NGFP up to planckian  energies, 
is to perform a renormalization group improvement of the classical lagrangian,  
a standard approach  in QED and QCD \cite{vacpol,dire}. 
For instance, for the quark binding problem in QCD \cite{adler2}
the leading contributions to the effective QCD lagrangian ${\cal L}^{\rm QCD}_{\rm eff}$  is encoded in the well known {\it leading-log} model,
which can be obtained by means of  a renormalization group improvement of the standard classical lagrangian, 
${\cal L}^{\rm QCD}_{\rm eff} = {{\cal F}}/{2 g^2_{\rm run}}$ where 
$g^2_{\rm run}={g^2(\mu^2)}/[{1+\frac{1}{4} \; b \; g^2(\mu^2) \log \left ({{\cal F}}/{\mu^4} \right )}]$, 
${\cal F}=-\frac{1}{2}(\partial_\mu A_\nu^a - \partial_\nu A_\mu^a+f^{abc}A_{\mu b} A_{\nu c})^2$, 
$\mu$ is an infrared subtraction scale and $b$ is the usual one-loop renormalization constant, dependent on the number of flavours. 
In this case the renormalization group is applied to the field-strength dependence rather than the usual $k^2$-dependence of the running coupling constant  
\cite{adler2,gross,masa,pato}.  A non-trivial confirmation of the validity of this approach in QCD can be found 
in the recent work of \cite{newlog} where it was shown
that the  leading-log model is in qualitative agreement with the results of an explicit 
evaluation of the full effective action obtained with the functional RG. 

Let us then start with  the standard classical EH Lagrangian
\be
{\cal L}^{\rm EH}=\frac{1}{16\pi G}(R-2\Lambda)
\label{eh}
\ee
and replace the coupling constant $G$ and $\Lambda$ in (\ref{eh}) with the running counterparts $G\rightarrow G(k)$ and 
$\Lambda \rightarrow \Lambda(k)$ with  $k^2 \propto R$.  
Although at this point it would be possible to consider more complicated tensor structure
of the curvature,  it has been shown in \cite{ntv} that,
at variance with Lagrangians dependent on non-linear function of the scalar Ricci curvature alone,
non-trivial vacua of higher derivative theory  dependent on non-linear functions of the square of the Ricci tensor or of the Riemann tensor, 
carry  additional 
ghost-like excitations which we will not be considered in the following discussion.
%

In particular, in our case  $G(k)=g(k)/k^2$ and $\Lambda(k)=\lambda(k) k^2$, 
being $g(k)$ and $\lambda(k)$ the dimensionless running Newton and Cosmological constant, respectively.
Since we do  not have an explicit solution for the $\beta$-functions at our disposal in general, it turns out to be  more convenient to use the information we know from the linearized flow around the NGFP.  In particular in 
\cite{2002PhRvD..66b5026L,2002CQGra..19..483L} it has been shown that the approach to the fixed point  is characterized by a pair of complex conjugate critical exponents  $\theta_1 = \theta^\star_2$ with positive real part $\theta'$ and imaginary parts $\pm \theta''$. Introducing $t = \ln (k/k_0)$ the general solution to the linearized flow equation reads
\ba\label{linear}
&&(\lambda, g)^{\mathbf T}=(\lambda_\ast, g_\ast)^{\mathbf T} 
+2 \{  [ {\rm Re}C \cos(\theta'' t) + {\rm Im}C  \sin(\theta'' t)] {\rm Re} \; V \nonumber\\
&&+[ {\rm Re}C \cos(\theta'' t) - {\rm Im}C \sin(\theta'' t)] {\rm Im} \; V \} \; e^{-\theta' t}
\label{linear}
\ea
where $C$ is an arbitrary complex number and $V$ is the right-eigenvector of the stability matrix (with eigenvalues $-\theta_1=-\theta^\star_2$)
around the NGFP in the EH truncation. Due to the fact that $\theta'>0$ all the trajectories converge to the fixed point $(\lambda_\ast, g_\ast)$
as $t\rightarrow \infty$. The imaginary part $\theta''$ plays no role in the stability but it influences the trajectories which spiral into the fixed point 
as $k\rightarrow \infty$.
We can now substitute the solution (\ref{linear}) in (\ref{eh}) obtaining the following {effective} lagrangian
\be
{\cal L}_{\rm eff}^{\rm QEG}(R) = R^2+b R^2 \cos \left [ \alpha \log \left ( \frac{R}{\mu} \right ) \right ]\; \left ( \frac{R}{\mu}\right )^\beta 
\label{impeh}
\ee
where the freedom in choosing the constant $C$ has been used in order to set  the coefficient of the sine term to zero.
In performing the RG-improvement $k^2 \propto R$ has been assumed,  following the same idea used  in obtaining the {\it leading-log} QCD effective action.
In addition,  a global constant rescaling  in order to have the coefficient of $R^2$ equal to unity
in the limit $R\rightarrow \infty$ has been performed: this is always possible as we do not consider the inclusion of matter fields.
In this notation, $\alpha=\theta''/2$, $\beta=-\theta'<0$ and $\mu=k_0^2$ is the infrared momentum scale which 
defines the crossover region between the gaussian-fixed point, and the NGFP.  
The constant $b$ could be calculated in principle 
from all the constants appearing in Eq.(\ref{linear}) and the fixed point values  $\lambda_\ast$ and $g_\ast$ but its precise value 
is not important in the following discussion. 

It is interesting to note that this particular a functional dependences of the {\it log-periodic} type as shown in Eq.(\ref{impeh})
characterizes a wide variety of physical systems like for instance the deviations from the pure Kolmogorov scaling in hydrodynamical turbulence
due to intermittence, or fractal structures.
The equation of motion for a generic action of the type (\ref{impeh}) are then obtained from 
\begin{eqnarray}
\label{lhs}
&&-\frac{1}{\sqrt{|g|}}\frac{\delta S}{\delta g_{\mu\nu}}
=\frac{d {\cal L}}{d R}\; R_{\mn}-\frac{1}{2}{\cal L}\; g_{\mn}-\nabla_\mu\nabla_\nu \frac{d {\cal L}}{d R} +g_{\mu\nu}\nabla^\rho\nabla_\rho \frac{d {\cal L}}{d R} = 0
\end{eqnarray}
being
\be
\label{action}
\S=\int d^4x\,\sqrt{|g|}\,{\cal L}(R)
\end{equation}
a generic gravitational action. 

Let us now investigate the physical content of this effective lagrangian in the context of early universe, 
by considering a spatially flat Friedmann-Robertson-Walker metric in vacuum.
In a FRW cosmology with scale factor $a(t)$ we can write both the Einstein tensor $G_{\mn}$ and the Ricci tensor
$R_{\mn}$ in terms of the Hubble rate $H(t)=\dot{a}(t)/a(t)$. In particular, 
the $(tt)$-component and (minus) the trace of (\ref{lhs}) become
\begin{eqnarray}
\label{Adef}
&&{\cal A}(H)=-3(\dot H+H^2) \frac{d {\cal L}}{d R}+ 3H{\frac{d {\dot {\cal L}}}{d R}}+\frac{1}{2}{\cal L}\\
&&{\cal B}(H)=-6(\dot H+2H^2)\frac{d {\cal L}}{d R}+2{\cal L}+3{\frac{d {\ddot {\cal L}}}{d R}}   + 9H {\frac{d {\dot {\cal L}}}{d R}} \,.
\label{Bdef}
\end{eqnarray} 
For the following analysis, instead of using directly Eq.(\ref{Adef}) it is more convenient 
to eliminate the ${\dddot H}$ term generated by (\ref{Bdef}) using (\ref{Adef}) in order to obtain
\ba
&&{\dot H}^2+6^\beta b \cos \left [ \alpha \ln \left ( \frac{6 (2 H^2 + {\dot H})}{\mu} \right ) \right ] \left (\frac{2 H^2 + {\dot H}}{\mu}\right)^\beta (2\beta H^4+(4\alpha^2-6\nonumber\\[2mm]
&&-\beta(9+4\beta))H^2 {\dot H} +(1+\beta){\dot H}^2+(\alpha-(1+\beta)(2+\beta))H {\ddot H}) \nonumber\\[2mm]
&&=2 H (3 H {\dot H} +{\ddot H})+6^\beta b \alpha \sin \left [ \alpha \ln \left ( \frac{6 (2 H^2 + {\dot H})}{\mu} \right ) \right ] \left (\frac{2 H^2 + {\dot H}}{\mu}\right)^\beta\nonumber\\[2mm]
&&(2H^4-(9+8\beta)H^2 {\dot H} +{\dot H}^2 - (3+2\beta) H {\ddot H})
\label{eqtf}
\ea
One is interested in de Sitter solutions, so that $H={\bar H}=const$ so that Eq.(\ref{eqtf}) yields
\be
{\bar H}= \sqrt{\frac{\mu}{12}}\; \exp \; \left [\frac{1}{2\alpha}  \left (\; {\tan}^{-1} \; \frac{\beta}{\alpha} + n \pi \right) \right ], \;\;\;\; n\in \mathbb{Z}
\label{solde}
\ee
which represent a countable number of de Sitter vacua all labeled by $n$. 
It is important to remark that this non-trivial structure of the solution is a consequence of the ``spiral" evolution around the NGFP.
As discussed by Weinberg \cite{weinberginflation}, the relevant question is if these solutions are unstable with characteristic growth time  
$\gg 1/{\bar H}$  so that inflation comes to an end after a large enough e-folds number $\approx 1/\xi$. 
In order to address this question it is convenient to write 
\be
H(t)={\bar H}+\delta H(t)
\ee 
and linearize Eq.(\ref{eqtf}) around the solutions (\ref{solde}) with $\delta H (t) = \exp(\xi {\bar H} t)$. 
After some manipulations, it is possible to obtain the following stability equation
\be
\xi^2+\xi \; 3 \,e^{\frac{n\pi}{2\alpha}}+A=0
\label{stab}
\ee
where
\be\label{12}
A=-\frac{4 \alpha  b (-1)^n \left(\alpha ^2+\beta ^2\right) e^{\frac{\beta  \tan ^{-1}\left(\frac{\beta }{\alpha
   }\right)+\pi  (\beta +1) n}{\alpha }}}{\alpha  b (-1)^n \left(\alpha ^2+\beta ^2-2\right) e^{\frac{\beta  \left(\tan
   ^{-1}\left(\frac{\beta }{\alpha }\right)+\pi  n\right)}{\alpha }}-2 \sqrt{\alpha ^2+\beta ^2}}
\ee
The interesting result of this discussion is that the stability of these inflationary solutions {does not depend on the mass scale  ${\mu}$}: only the real and imaginary part of the critical exponent and the point in the 
$\lambda$-$g$-plane, monitored by the constant $b$, determine the stability of the solution. 
As discussed in \cite{2002CQGra..19..483L}, the scheme dependence of the critical exponents turns out to be very limited, as 
$\theta'$ and $\theta''$ assume values in the ranges $2.1 < \theta' < 3.4$ and $3.1 < \theta'' < 4.3$, respectively for various cutoff functions.    
For positive values of the integer $n$ the constant $A$ decays exponentially to zero because $\beta<0$, and one
is left only with a negative root, which implies stability, for any value of $b$. 
The situation is different for negative values of $n$ because the exponents in (\ref{12}) are very large, the 
$-2 \sqrt{\alpha ^2+\beta ^2}$ term in the denominator can be neglected and the $b$-dependence cancels out:
now $A$ is always negative and one root is unstable. 
In this case $A$ can be approximated with
\be
A\approx -\frac{4 \left(\alpha ^2+\beta ^2\right) e^{\frac{\pi  n}{\alpha }}}{\alpha ^2+\beta ^2-2}
\ee
and the unstable root is 
\be
\xi=\frac{\left(\sqrt{\left(\alpha ^2+\beta ^2-2\right) \left(25 \alpha ^2+25 \beta ^2-18\right)}-3 \alpha ^2-3 \beta
   ^2+6\right) e^{\frac{\pi  n}{2 \alpha }}}{2 \left(\alpha ^2+\beta ^2-2\right)}
\ee
The prefactor in front of the exponential is always of the order unity and positive for $\theta'$ and $\theta''$ in the allowed range
as $\alpha=\theta''/2$ and $\beta=-\theta'$ and therefore  it is always possible to produce enough e-folds of inflation 
for $n$ negative enough. At last  we find
\be
1/\xi \approx e^{-n\pi/\theta''} 
\label{efolds}
\ee
for the number of e-folds. For instance for $n=-3$,  $1/\xi \approx 17$   and  for $n=-4$ one gets $1/\xi \approx 49$
while for $n=-5$, $1/\xi \approx 140$. It should be stressed that this result is rather remarkable, because  
it only depends on one ``universal" quantity namely the imaginary part of the critical exponent which  characterizes 
the flow around the NGFP. This is reassuring because otherwise a strong cutoff dependence  
in (\ref{efolds}) would have signaled that important terms in the truncations still needed to be
considered in the {\it cos-log} model of Eq.(\ref{impeh}). 

The effective action presented in this work has some attracting features, which could be useful in discussing the cosmological consequences of an asymptotically safe
gravity near the planck scale. It reproduces the expected leading $R^2$ behavior of the lagrangian for large curvature near the NGFP due to the fact that $G \sim 1/k^2$ at the NGFP, but it also embodies the important information provided by the linearized RG flow which ``spirals" towards the NGFP. 

One limitation of our result is that it has been obtained within the EH truncation. However, it is not difficult to realize that the structure of the linearized flow 
near the NGFP can always be recast in an effective lagrangian of the {\it log-periodic} type. 
To illustrate this point, let us consider a quadratic lagrangian of the type introduced in \cite{2002PhRvD..66b5026L}
\be
\label{in2}
{\cal S}=\int d^4 x\,\sqrt{|g|} \left \{ \frac{1}{16\pi G} (R-2\Lambda)  - \beta R^2 \right \}
\ee
The linearized flow in the vicinity of NGFP is also governed by a pair of complex conjugate critical
exponents $\theta_1=\theta'+{\rm i}\theta''=\theta_2^*$ with $\theta'>0$ and
a single real, positive critical exponent $\theta_3>0$. It may be expressed as
\begin{eqnarray}
\label{p2.6}
\left(\lambda_k,g_k,\beta_k\right)^{\bf T}
&=&\left(\lambda_*,g_*,\beta_*\right)^{\bf T}
+2\Big\{\left[{\rm Re}\,C\,\cos\left(\theta''\,t\right)
+{\rm Im}\,C\,\sin\left(\theta''\,t\right)\right]
{\rm Re}\,V\\
& &+\left[{\rm Re}\,C\,\sin\left(\theta''\,t\right)-{\rm Im}\,C
\,\cos\left(\theta''\,t\right)\right]{\rm Im}\,V\Big\}\,e^{-\theta' t}
+C_3 V^3\,e^{-\theta_3 t}\nonumber
\end{eqnarray}
with arbitrary complex $C\equiv C_1=(C_2)^*$ and arbitrary real $C_3$, and
with $V\equiv V^1=(V^2)^*$ and $V^3$ the right-eigenvectors of the stability
matrix  with eigenvalues $-\theta_1=
-\theta_2^*$ and $-\theta_3$, respectively. The conditions for UV 
stability, $\theta'>0$ and $\theta_3>0$, are satisfied for all 
cutoffs, but since  $\theta_3\gg \theta'$,  it is difficult to imagine that the
renormalized flow is strongly affected by the presence of the $R^2$ term in the lagrangian.
On the other hand, we know from the work of \cite{cpr2}
that the values of $\theta'$ and $\theta''$ are 
not significantly changed by the presence of higher order polynomial up to $R^8$ in the action. 

It should be remarked that our discussion eventually breaks down in the IR, around the gaussian-fixed point, where possibly a new set
of IR relevant operators can show up. 

\end{document}